\documentstyle[twoside,fleqn,espcrc2,epsfig]{article}

\newcommand {\sax} {{\it BeppoSAX }}
\newcommand {\etal} {et~al. }

\newcommand{\AmS}{{\protect\the\textfont2
  A\kern-.1667em\lower.5ex\hbox{M}\kern-.125emS}}

\title{X-Ray Luminosity and Spectral Variability in the TEV BL Lac 1ES2344+514
       }

\author{P. Giommi, \address{\sax Science Data Center,Rome, Italy}
       P. Padovani \address{Space Telescope Science Institute, Baltimore, USA}$^,$
                   \address{Affiliated to the Astrophysics Division, Space Science Department, ESA}$^,$
                  \address{On leave from Dipartimento di Fisica, II Universit\'a di Roma, Italy}
        and 
       E. Perlman$^{\rm b}$ 
       }
       
\begin{document}

\begin{abstract}

The results of a series of five \sax observations of the TeV BL Lac 
object 1ES2344+514 are briefly presented. 
Large amplitude luminosity variability, associated to impressive spectral 
changes in the hard X-rays, have been found. The shape of the lightcurve 
depends on energy, with the flare starting and ending in the hard band, 
but with maximum intensity possibly reached earlier in the soft X-rays.
The luminosity and spectral changes may be due to a shift of 
the peak of the synchrotron emission from the soft X-rays to the hard 
X-ray band similar to that detected during \sax observations of MKN 501.
\end{abstract}

\maketitle

\section{Introduction}

BL Lac objects are a peculiar type of radio loud AGN emitting highly
variable non-thermal radiation over an extremely wide energy range from 
radio waves to TeV energies.
Synchrotron emission followed by Inverse Compton scattering is generally 
thought to be the mechanism responsible for the production of non-thermal 
radiation over such a wide energy range (e.g. Bregman \etal 1994). 
Relativistic beaming is necessary to explain some of the extreme properties 
of these objects such as rapid variability and superluminal motion (Urry \& Padovani 1995). 1ES2344+514 is an HBL BL Lac (Padovani \& Giommi 1995) that 
is an object where the synchrotron component dominates the spectrum up 
to very high energies. 
To date only three BL Lacertae objects (all of them HBLs) have been 
detected at TeV energies (Catanese et al. 1997): MKN 501, MKN 421 and 
1ES2344+514.

We briefly present here the main results of a series of broad-band 
(0.1-200 keV) X-ray observations of 1ES2344+514 carried out with 
the \sax satellite (Boella \etal 1997a).

\begin{figure}[h*]
\epsfig{figure=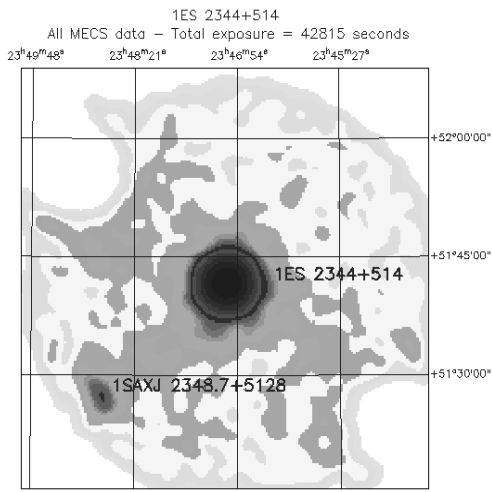, height=7.0cm, width=7.0cm}
\caption{The MECS X-Ray image of the field centered on 1ES2344+514. 
The bright source at the center is the BL Lac, the faint serendipitous 
source (1SAXJ2348.7+5128) to the bottom left part of the image coincides 
with the dwarf nova V630 CAS}
\label{fig1}
\end{figure}

\section{Observations}

The \sax  satellite carries aboard four co-aligned 
X-ray experiments called NFI (Narrow Field Instruments) and two Wide-Field-Cameras (WFC) each pointing 90 degrees away from the NFI pointing direction.
The NFI include four X-ray telescopes with one Low Energy Concentrator Spectrometer (LECS, Parmar \etal 1997) sensitive in the 0.1-10 keV band, and three identical Medium Energy Concentrator Spectrometers (MECS, Boella 
\etal l997b) covering the 1.5-10. keV band. 
At higher energy the NFI also include two collimated experiments (PDS, Frontera et al. 1997, and HPGSPC, Manzo et al. 1997) extending the \sax 
sensitivity band to approximately 200 keV.

The journal of the \sax NFI observations of 1ES2344+514 together with the count-rates in the LECS, MECS and PDS instruments is given in table 1.
 
\begin{table*}[hbt]
\setlength{\tabcolsep}{1.0pc}
\newlength{\digitwidth} \settowidth{\digitwidth}{\rm 0}
\catcode`?=\active \def?{\kern\digitwidth}
\caption{Log of the \sax observations 1ES 2344+514}
\label{tab:log}
\begin{tabular*}{\textwidth}{@{}l@{\extracolsep{\fill}}rccccc}
\hline
Observation & LECS & count-rate &  MECS & count-rate 
& PDS &count-rate \\
date& exp (s) & $ct~s^{-1}$ &exp (s) & $ct~s^{-1}$ &exp (s) 
& $ct~s^{-1}$\\
\hline
3-DEC-1996 &  4719&$0.19\pm 0.01$ & 13109& $0.42\pm 0.01$& 6375& $<0.3$ \\
4-DEC-1996 &  5264&$0.22\pm 0.01$ & 13300& $0.42\pm 0.01$&6518&$0.27\pm0.09$  \\
5-DEC-1996 &  ~986&$0.31\pm 0.02$ & ~2547& $0.54\pm 0.02$&1228& $<0.6$ \\
7-DEC-1996 &  5563&$0.35\pm 0.01$ & 14069&$0.82\pm 0.01$ &6757&$0.32\pm0.09$\\
11-DEC-1996&  2992&$0.22\pm 0.01$ & 13062&$0.53\pm 0.01$ &6113& $<0.3$ \\
\hline
\end{tabular*}
\end{table*}

\section{Data analysis}

The analysis of the LECS and MECS data was carried out using the screened 
event files made available through the \sax Science Data Center (SDC) on-line 
archive (Giommi \& Fiore 1997). The XIMAGE analysis software 
(Giommi \etal 1991) (upgraded to support \sax data) and the XSPEC package 
were used for image and spectral analysis. 
The MECS 2-10 keV image, accumulated using data from all observation is 
shown in figure 1. 1ES2344+514 is the bright source in the center, 
while the fainter serendipitous source visible at the bottom left of the 
image (1SAXJ2348.7+5128) is the dwarf nova V630 CAS.
The count rates of 1E2344+514 are reported in table 1 where it can be seen 
that intensity variations of approximately a factor 2 are present in all 
experiments. 
The count rate of 1SAX J2348.7+5128 did not show any significant variation.

Spectra of 1E2344+514 were accumulated for each observation, 
with 8.5 and 4 arcmin extraction radii for the 
LECS and MECS instruments respectively. 
The PDS spectra were instead taken directly from the \sax on-line archive
(Giommi \& Fiore 1997) which is based on the results of the
supervised standard analysis. 
 The LECS data have been fitted in the $0.1-4$ keV range, 
while MECS data were fitted over the full $1.8-10.5$ keV range.
Spectral analysis was performed assuming both a simple and a broken 
power law model with photoelectric absorption fixed to the Galactic 
value of $1.6\times 10^{21}~cm^{-2}.$

We find that a simple power law is in general an acceptable 
representation of the data, although in some cases the best fit 
requires some absorption in excess of the Galactic value. This is 
probably the symptom of a spectral flattening in the soft X-rays (i.e. 
the spectrum is convex), as is often seen in HBL  
BL Lacs (e.g. Padovani \& Giommi 1996). This interpretation is fully
supported by the results of the broken power law fits which indicate 
steeper slopes in the hard band. 
The spectral slope varies with flux  from a value of around 1.0 
(energy index) when the source was faint to a harder value 
of about 0.77 when 1ES2344+514 was at maximum intensity. 
The largest spectral change occurred above 8-10 keV, mostly 
in the PDS band. 
This is illustrated in figures 2 and 3 where  the LECS, MECS and PDS 
data are plotted together with the power law best fit, for 
the observations made on December 3 and December 7 when the source was 
faint and brightest respectively. 
A more effective representation of the spectral change is shown in 
figure 4 where the \sax data taken on December 3, 4 and 7 are plotted 
in a $\nu f(\nu)~vs~\nu$ energy distribution representation. 

\begin{figure}[ht]
\epsfig{figure=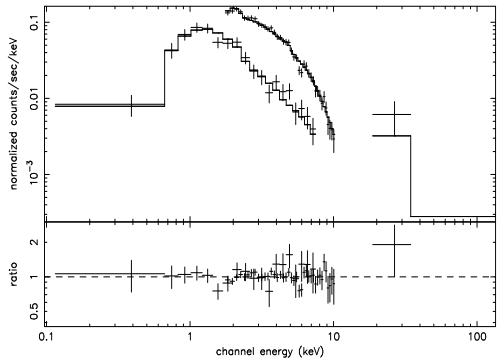, height=7.5cm, width=6.0cm, angle=-90}
\caption{The LECS, MECS and PDS spectrum of 1ES2344+514 in the faint 
state. The object is barely detected in the PDS instrument up to 
20-30 keV. No detection is present at higher energy.}
\label{fig2}
\end{figure}

\begin{figure}[ht]
\epsfig{figure=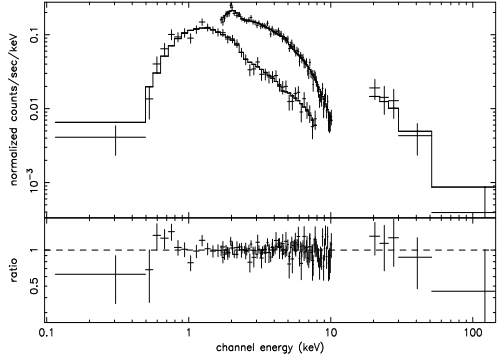, height=7.5cm, width=6.0cm, angle=-90}
\caption{The LECS, MECS and PDS spectral data for 1ES2344+514 taken 
on December 7 when the source was brightest.}
\label{fig3}
\end{figure}

\section{X-Ray flux and spectral variability}

Figure 5 shows the X-ray lightcurve of 1ES 2344+514 in three 
bands : 0.1-2 keV, 3-5 keV, and 5-10 keV.
A visual inspection of this figure shows that the brightening 
starts at least one day earlier in the hardest band where it also 
lasts longer. Maximum intensity seems to be reached earlier in 
the softest band.  

The LECS, MECS and PDS data for the observations of December 3, 4 
and 7 have been combined in the spectral energy distribution 
($\nu f(\nu)~vs~\nu $) shown in figure 4 where a spectacular spectral 
variability is apparent in the highest energy bins. 
The December 5 and 11 data are not shown on the plot to avoid 
confusion but are fully consistent with the general trend of a strong 
spectral hardening with intensity apparent from the figure.

\begin{figure}[ht]
\epsfig{figure=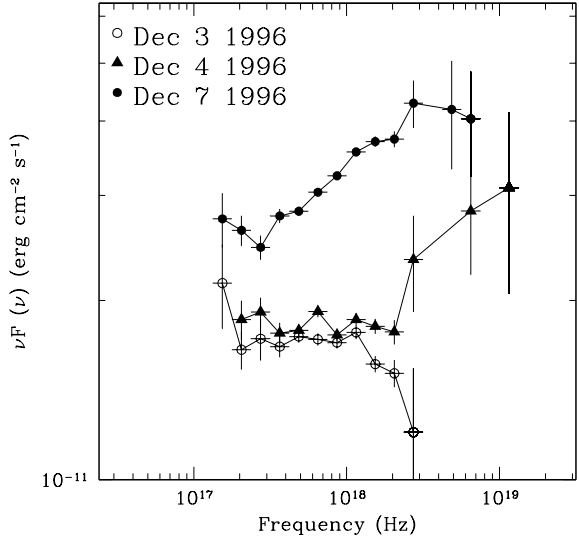, height=8.cm, width=8.0cm}
\caption{The LECS, MECS and PDS spectral data plotted as 
$\nu f(\nu)~vs~\nu $. Very large spectral variations, especially at 
high energy, are clearly visible.
}
\label{fig5}
\end{figure}

\begin{figure}[ht]
\epsfig{figure=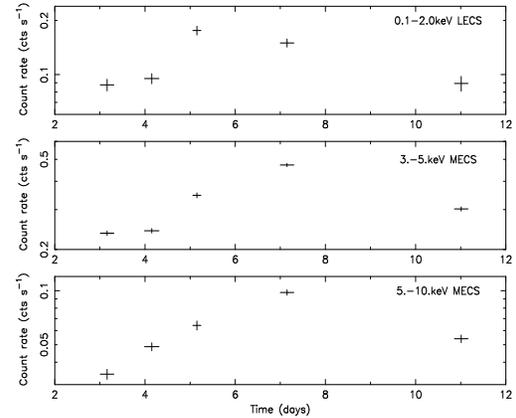, height=7.5cm, width=6.0cm, angle=-90}
\caption{The lightcurve of 1E2344+514 is three X-ray energy bands}
\label{fig4}
\end{figure}

\section{Discussion}

During a series of five \sax observations 1ES 2344+514 underwent a flare 
which caused its luminosity to double in the 0.1-10 keV band and to increase
of an even larger amount in the PDS band.
The X-ray spectral shape of 1ES2344+514 varied with intensity in a way 
that is typical of HBL BL Lacs, namely the spectrum hardens when the 
source brightens (e.g. Giommi et al. 1990).
The variable X-ray spectral energy distribution of 1ES2344+514 
($\nu f(\nu)~vs~\nu $) is consistent with an interpretation where the 
flux and the spectral variations are due to the onset of a variable hard 
component that started dominating the X-ray flux above 5 keV during 
the second observation and that reached maximum intensity on 
December 7, when it dominated the entire X-ray band. 
Figure 4 shows that the peak of the synchrotron power
(that is where the spectral slope is 0 in $\nu f(\nu)~-~\nu $ space)
during the low state was at frequencies of a few times $10^{17}$ while 
when the intensity was maximum the peak shifted to $\approx 3-4\times 10^{18}$
(20-30 keV). 

A similar spectral variability, but on an even
larger scale, was detected in a recent observation of
the X-ray bright and TeV detected HBL BL Lac MKN 501 (Pian et al. 1998).
Large shifts of the synchrotron peak energy (implying very large 
changes in the bolometric luminosity) might therefore be relatively 
frequent in the hard X-ray band for HBL BL Lacs. 
Similar behavior should be expected in LBL BL Lacs in the optical/UV band. 
As for the case of MKN501 (Pian et al. 1998) the results reported
here demonstrate that most of the power emitted in BL Lacs
may be in the hard X-Rays ($ \ge 10 $ keV) an energy region
poorly explored by previous satellites. 

\end{document}